\documentstyle[12pt]{article}

\def\be{\begin{equation}}
\def\ee{\end{equation}}
\def\ben{$$}
\def\een{$$}
\def\ba{\begin{array}{c}}

\def\ea{\end{array}}
\def\p{\partial}
\begin{document}

\titlepage
\begin{center}
.

\vspace{2cm}

{\Large \bf Exactly solvable three-body Calogero-type model

with translucent two-body barriers }\end{center}

\vspace{10mm}

\begin{center}
Miloslav Znojil and Milo\v{s} Tater

\vspace{3mm}

odd\v{e}len\'{\i} teoretick\'{e} fyziky,\\
 \'{U}stav jadern\'e
fyziky AV \v{C}R, 250 68 \v{R}e\v{z},
 Czech Republic\footnote{
 e-mail: znojil@ujf.cas.cz and
tater@ujf.cas.cz }\\

\end{center}

\vspace{5mm}

\section*{Abstract}

A new exactly solvable alternative to the Calogero three-particle
model is proposed. Sharing its confining long-range part, it
contains the mere zero-range two-particle barriers. Their
penetrability gives rise to a tunneling, tunable via their three
independent strengths. Their variability can control the removal
of the degeneracy of the energy levels in an innovative,
non-perturbative manner.

\vspace{10mm}

PACS 21.45.+v 03.65.Ge 02.60.Lj

\newpage

\section{Introduction}


In nuclear, atomic, molecular and statistical physics, the
enormous popularity of the Calogero's solvable models
\cite{Calogero} reflects a fairly realistic form of their separate
two-body interactions. They combine a long-range quadratic
attraction with a strong short-range repulsion. In its full
generality the model binds an arbitary number $N$ of particles
moving along a straight line and its exact solvability is a
consequence of certain deep symmetries of its partial differential
Schr\"{o}dinger equation \cite{Turbiner}.

For the sake of brevity, let us pay attention just to the
Calogero's first nontrivial three-body Hamiltonian
 \ben
 H^{(Cal)} = -
 \frac{\hbar}{2m}\,
 \left[
\frac{\p^2}{\p{x_1}^2} + \frac{\p^2}{\p{x_2}^2} +
\frac{\p^2}{\p{x_3}^2} \right ] + \ \ \ \ \ \ \ \ \ \ \
 \ \ \ \ \ \ \ \ \ \een \ben
 + \frac{1}{8}\,\omega^2   \, \left [
(x_1-x_2)^2
 +(x_2-x_3)^2
 +(x_3-x_1)^2 \right ] +
\ \ \ \ \ \ \ \ \ \ \
 \een
 \be
 \ \ \ \ \ \ \ \ \ \ \ \ \ \
 \
 + \left [
 \frac{g_1}{(x_2-x_3)^2}
  +\frac{g_2}{(x_3-x_1)^2}+
\frac{g_3}{ (x_1-x_2)^2}
 \right ]\ .
 \label{Calo}
  \ee
We may immediately see that from a purist's point of view, its
two-body barriers are not penetrable \cite{Landau} and, in this
sense, they do not admit any mutual exchange of particles. For
this reason we proposed recently a solvable modification of this
model where a tunneling has been rendered possible at an expense
of a loss of the Hermiticity of the Hamiltonian $H^{(Cal)}$
\cite{Milos}. We employed a complexification  based on the shift
of the coordinates $x^{(real)} \to x^{(complex)}(t) = t -
i\,\varepsilon(t^2)$. This replaced the Hermiticity of the
Hamiltonian by its mere commutativity with the product of parity
${\cal P}$ and complex conjugation ${\cal T}$ \cite{BB}. We have
shown that the spectrum remained real in a way attributed to the
${\cal PT}$ symmetry in the related literature \cite{BBjmp}.

In the latter innovative few-body implementation of the idea of
the ${\cal PT}$ symmetrization, the Calogero's strongly singular
real barriers $1/x^2$ with $ x = x_j-x_k$ were all replaced by the
complex and $\varepsilon-$dependent expressions
 \be
 \frac{1}{(x-i\varepsilon)^2}
= \frac{1}{x^2+\varepsilon^2}
+ \frac{2i\varepsilon\,x}{(x^2+\varepsilon^2)^2}
 + {\cal O}(\varepsilon^2).
 \label{leading}
 \ee
In the present short note we intend to describe and analyze an
alternative scheme which would admit a tunneling.  In essence, we
shall start from the same leading-order formula (\ref{leading})
but succeed in returning to a Hermitian Hamiltonian.

\section{The new model}

\subsection{Inspiration}

Our new proposal has been inspired by formula (\ref{leading}) and
by its Hermitization
 \be
  \frac{1}{x^2} \to
  \frac{1}{x^2+\varepsilon^2} =
 \frac{1}{2i\varepsilon}\,
 \left [
 \frac{1}{x-i\varepsilon}
 -
 \frac{1}{x+i\varepsilon}
 \right ]
  = \frac{\pi}{\varepsilon} \tilde{\delta}_\varepsilon(x),
  \ \ \ \ \ \ \ \ \varepsilon > 0.
 \label{hlead}
 \ee
The limit $\varepsilon \to 0$ would then just reproduce the
Calogero's model, one of the specific and most inspiring features
of which lies in its separability in the three-body case.  This is
based on the re-parametrization of the coordinates $x_1-x_2 =
\sqrt{2}\,\rho \, \sin \phi$ with $\rho \in (0,\infty)$ and $\phi
\in (0, 2\pi)$ etc (cf. Figure~1). In the units $2m = \hbar =1$
and for the equal and non-negative coupling constants
$g_1=g_2=g_3=g\geq 0$ in eq. (\ref{Calo}), such a change of
variables reduces the Calogero's partial differential
Schr\"{o}dinger eqution to the mere ordinary Sturm -- Liouvillean
problem
 \be
 \left ( -\frac{d^2}{d \phi^2} + \frac{9\,g}{2\,\sin^23\,\phi}
 \right ) \psi(\phi) = \kappa^2\psi(\phi)
 \label{angular}
, \ \ \ \ \ \ \ \ \ \ \ \  \psi(0)=\psi(2\pi) = 0.
 \label{bca}
 \ee
It solution generates the well known (viz., Laguerre times Jacobi)
polynomial wave functions as well as the related equidistant
spectrum with a gap, $E_{n} \sim n + const$, $n = 0, 2,3,4,
\ldots$.

\subsection{Main idea}

Our attention is attracted by the tilded expression
$\tilde{\delta}_\varepsilon(x)$ in eq. (\ref{hlead}).  We shall
try to re-read it as an approximate $\varepsilon \approx 0$ form
of the well known Dirac delta function  $\tilde{\delta}_0(x)
={\delta}(x)$. Thus, picking up, say, the first part of the
two-body barrier in eq. (\ref{Calo}) in its regularized form
 \ben
\frac{g_1}{ (x_2-x_3)^2+\varepsilon^2}
 = f_1 \tilde{\delta}_\varepsilon(x_2-x_3) ,
\ \ \ \ \ \  \ \ \
f_1=\frac{\pi\,g_1}{\varepsilon}
,\ \ \ \ \ \  \ \ \
 \varepsilon > 0,
 \een
we may insert the appropriate definition $x_2-x_3 = \sqrt{2}\,\rho
\, \sin( \phi-4\pi/3)$ and get, at the very small $\varepsilon \to
0$,
 \ben
 f_1 \tilde{\delta}_\varepsilon(x_2-x_3)
\approx
\frac{ G_1}{\rho^2}\,
\left [
\tilde{\delta}_\varepsilon \left ( \phi- \frac{\pi}{3} \right )
+\tilde{\delta}_\varepsilon \left ( \phi- \frac{4\pi}{3} \right )
\right ]
,
\ \ \ \ \ \ \ \ \ \ G_1= \frac{f_1\rho}{\sqrt{2}}.
 \label{trial}
 \een
In place of (\ref{angular}) it
gives
 \ben
  -\frac{d^2}{d \phi^2}\psi(\phi)  +
G_1 \left [
 \tilde{\delta}_\varepsilon \left ( \phi- \frac{\pi}{3} \right )
+\tilde{\delta}_\varepsilon \left ( \phi- \frac{4\pi}{3} \right )
\right ]\, \psi(\phi)
+
\ \ \ \ \ \ \ \ \ \ \  \ \
\ \ \ \ \ \ \  \ \
\een
\ben
\ \ \ \ \ \ \
+
G_2 \left [
 \tilde{\delta}_\varepsilon \left ( \phi- \frac{2\pi}{3} \right )
+\tilde{\delta}_\varepsilon \left ( \phi- \frac{5\pi}{3} \right )
\right ]\, \psi(\phi)+
\ \ \ \ \ \ \
\een
\be
\ \ \ \ \ \ \ \ \ \ \ \ \ \
+
G_3 \left [
 \tilde{\delta}_\varepsilon \left ( \phi- \pi \right )
+\tilde{\delta}_\varepsilon \left ( \phi \right )
\right ]
\, \psi(\phi)
 = \kappa^2\,\psi(\phi),
 \label{gular}
 \ee
i.e., an approximative innovation of the Calogero's angular
Schr\"{o}dinger equation.

\subsection{Interpretation}

Due to the purely intuitive form of the above ``derivation" of eq.
(\ref{gular}), one has to be very careful in all questions related
to its possible physical as well as mathematical interpretation.
At the same time, the use of the delta-function-shaped potentials
is quite common in practice \cite{Fluegge} as it makes many
systems exactly solvable in the limit $\varepsilon \to 0$.

Let us start our further analysis of eq. (\ref{gular}) by picking
up some three constants $G_k> 0$. Then we re-define
$f_k=f_k(\rho)=\sqrt{2}G_k/\rho$ and
$g_k=\varepsilon\,f_k(\rho)/\pi$.  In the generic case with
$x_i-x_j \neq 0$, the limiting transition to $\varepsilon = 0$
makes all the three centrifugal-like forces vanish, $g_k \to 0$.
Such an observation is compatible with the philosophy of using
just the contact barriers in our angular Schr\"{o}dinger equation.

In the second step we have to re-analyze the role of the overall
singularity in the origin $\rho = 0$. In principle, we might admit
and consider many different singularities there \cite{Sukhatme}.
Here we shall simplify our life by the most straightforward
postulate that the point $\rho=0$ is just ``regular" (i.e.,
``ignored" by our Hamiltonian), and that all our radial wave
functions are simply vanishing in the origin. Under such an
assumption, the final $\rho \neq 0$ form of our new Hamiltonian $
H^{(new)}$ can be assigned a formal representation
 \ben
 H^{(new)} = -
 \frac{\hbar}{2m}\,
 \left[
\frac{\p^2}{\p{x_1}^2} + \frac{\p^2}{\p{x_2}^2} +
\frac{\p^2}{\p{x_3}^2} \right ] + \ \ \ \ \ \ \ \ \ \ \
 \ \ \ \ \ \ \ \ \ \een \ben
 + \frac{1}{8}\,\omega^2   \, \left [
(x_1-x_2)^2
 +(x_2-x_3)^2
 +(x_3-x_1)^2 \right ] +
 \een
 \be
 \ \ \ \ \ \ \ \ \ \ \ \ \ \
 \
 +
 {{\Omega}_1}{\delta(x_2-x_3)}+
 {{\Omega}_2}{\delta(x_3-x_1)}+
 {{\Omega}_3}{\delta(x_1-x_2)}
 \label{newer}
  \ee
where the strength of the two-body contact terms has the
three-body character and weakens with the distance of the detached
spectator particle,
 \ben
\Omega_1=\frac{\sqrt{3}\,G_1}{|x_1-x_2|}, \ \ \ \ \ \ \
\Omega_2=\frac{\sqrt{3}\,G_2}{|x_2-x_3|}, \ \ \ \ \ \ \
\Omega_3=\frac{\sqrt{3}\,G_3}{|x_3-x_1|}.
 \een
We may summarize that our Hamiltonian $ H^{(new)}$ represents a
new three-particle model which is exactly solvable.

\section{Solutions}


Our new angular Schr\"{o}dinger equation
 \ben
  -\frac{d^2}{d \phi^2}\psi_k(\phi)  +
G_1 \left [
 {\delta} \left ( \phi- \frac{\pi}{3} \right )
+{\delta} \left ( \phi- \frac{4\pi}{3} \right )
\right ]\, \psi_k(\phi)
+
\ \ \ \ \ \ \ \ \ \ \  \ \
\ \ \ \ \ \ \  \ \
\een
\ben
\ \ \ \ \ \ \
+
G_2 \left [
 {\delta} \left ( \phi- \frac{2\pi}{3} \right )
+{\delta} \left ( \phi- \frac{5\pi}{3} \right )
\right ]\, \psi_k(\phi)+
\ \ \ \ \ \ \
\een
\be
\ \ \ \ \ \ \ \ \ \ \ \ \ \
+
G_3 \left [
 {\delta} \left ( \phi- \pi \right )
+{\delta} \left ( \phi \right )
\right ]
\, \psi(\phi)
 = \kappa_k^2\,\psi_k(\phi)
 \label{lar}
 \ee
describes the $\phi-$dependent part of the three-body wave
function and determines its energies via the Calogero-type
formula,
 \be
E=E_{m,k}=
\sqrt{\frac{3}{2}} \, \omega\,(2m+1 +\kappa_k), \ \ \ \ \ m, k =
0, 1, \ldots \ .
 \ee
Within the open subintervals of $\phi \in {\cal D}_j =
([j-1]\,\pi/3,j\,\pi/3)$ we have a free motion
 \ben
 \psi(\phi) =
 \psi_j(\phi) =
A_j\sin \{ \kappa\,[\phi-(j-1)\,\pi/3] \} +
B_j\cos \{ \kappa\,[\phi-(j-1)\,\pi/3] \}.
 \een
On the boundaries we have to match the separate local wave
functions using the standard rules
 \ben
 \psi_j
 \left (\frac{j\,\pi}{3}
 \right ) = B_{j+1}, \ \ \ \ \ \ \
\p_\phi  \psi_j \left (\frac{j\,\pi}{3}
 \right )  = \kappa\,[A_{j+1} +\beta_j(\kappa)B_{j+1}], \ \
\ \ \ \ \ \ \ \ j = 1, 2, \ldots, 6
 \een
with $\beta_j(\kappa)=\beta_{j+3}(\kappa) =G_j/\kappa$ and $j = 1,
2$ or $3$. The first five applications of these rules define all
the wave functions $ \psi_j(\phi)$ in terms of the initial one,
say, $\psi_1(\phi)$. The sixth step becomes a selfconsistent
matching which guarantees that the global wave function remains
single-valued. In terms of the two auxiliary matrices
 \ben
R=R(\kappa)=
\left (
\begin{array}{cc}
\cos (\kappa\,\pi/3)& \sin (\kappa\,\pi/3)\\ -\sin
(\kappa\,\pi/3)& \cos (\kappa\,\pi/3) \ea \right ), \ \ \ \ \ \ \
\ L
 =
\left (
\begin{array}{cc}
0&0\\
1&0\ea
\right )
 \een
our matching conditions factorize in the two independent
two-dimensional forms
 \ben
( {\cal U} \pm I )\, \left ( \ba B_1\\A_1 \ea \right ) = 0
 \een
where $ {\cal U} = {\cal U}(\kappa) = (I+\beta_3L)R\,
(I+\beta_2L)R\, (I+\beta_1L)R$. The pertaining two alternative
two-dimensional secular equations
 \be
 \det [ {\cal U}(\kappa^{(\pm)}) \pm I] = 0
 \label{secular}
 \ee
define all the roots $\kappa^{(\sigma)}_k\geq 0$ with the sign
ambiguity $\sigma=\pm 1$ and the angular quantum number $k=0, 1,
\ldots$ in implicit manner.

\subsection{Toy model with the single barrier}

Let us fix $G_1=G_2=0$ and vary the strength $G = G_3$. With
$\beta=G/\kappa^{(\sigma)}$ in the matrix
 \ben
  {\cal U}^{(toy)}(\kappa) = (I+\beta\,L) \cdot
\left (
\begin{array}{cc}
\cos \kappa\,\pi
&
\sin \kappa\,\pi\\
-\sin \kappa\,\pi&
\cos \kappa\,\pi
\ea
\right )
\een
our determinantal secular equation reads
 \ben
\cos \kappa\,\pi + \sigma + \frac{G\,\sin \kappa\,\pi}{2\kappa}=0.
 \een
At $\sigma=-1$ and any integer shift of $\kappa^{(-)}=2k+\delta$
with $k = 0, 1, \ldots$, this secular equation gives us two roots
$\delta(a),\delta({b}) \in [0,1)$. The smaller one is trivial,
$\delta(a)=0$. The second one is uniquely specified by the
implicit formula
 \ben
\tan \frac{\pi\delta(b)}{2} = \frac{G}{4k+2\delta(b)}.
 \een
Also for $\sigma=+1$ and $\kappa^{(+)}=2k+1+\delta$ we get
$\delta(a)=0$ and the similar relation
 \ben
\tan \frac{\pi\delta(b)}{2} = \frac{G}{4k+2+2\delta(b)}.
 \een
In the limit $G\to 0$ of the vanishing barrier we get the standard
square-well solutions with the correct degeneracy
$\delta(b)\to\delta(a)$ at every $k$.

The sign of the coupling depends on our choice, $G\in
(-\infty,\infty)$. This provides an interesting counterpart to the
Calogero model where the barrier cannot be too attractive
\cite{0102034}. In the limit $G \to \infty$ of the very strong
repulsion, one returns to the Calogero-like case characterized by
the impenetrability of the barriers.

\subsection{Three equal barriers}

When we choose $G_1=G_2=G_3=G$ and abbreviate $ (I+\beta\,L)R=
{\cal R}(\kappa) $ with $\beta=G/\kappa$, our secular equation
(\ref{secular}) may be factorized into the two separate conditions
at both the quasi-parities $\sigma = \pm 1$,
 \ben
 \det [{\cal R}(\kappa) +\sigma I]=0,
\ \ \ \ \ \ \ \ \
 \det [{\cal R}^2(\tilde{\kappa})-\sigma {\cal R}
(\tilde{\kappa})  + I]=0.
 \een
The first one parallels our preceding toy model and its solution
is immediate,
 \ben
\kappa(a) = 3N, \ \ \ \
\kappa(b) = 3N + 3\delta(b), \ \ \ \ \ N = 0, 1, \ldots\ .
 \een
The implicit definition of the pertaining shifts $\lambda (b)\in
(0,1)$ has almost the same form as above,
 \ben
\tan \frac{\pi\delta(b)}{2} = \frac{G}{6N+6\delta(b)}.
 \een
In the tilded case, the identity $ \det ({\cal R}^2-\sigma {\cal
R} + I)= ({\sigma-\rm tr}\,{\cal R})^2$ holds as long as $\det
{\cal R}=1$. This reduces the second secular condition to the
trigonometric equation
 \ben
2 \, \cos \frac{\tilde{\kappa}\,\pi}{3}+
\frac{G}{\tilde{\kappa}}\, \sin \frac{\tilde{\kappa}\,\pi}{3}=
 \sigma .
 \een
Its solutions must be sought numerically, giving $\tilde{\kappa}=
1.367840720, \ 2.199769250, \ldots$ at $G=1$ (cf. Figure~2).

\section{Summary}

One of the most striking features of the model of Calogero is the
fully impenetrable character of its two-body repulsive barriers.
They divide the phase space in the six independent subdomains (cf.
Figure~1). In a way, this absolute impenetrability of the barriers
is in its effect responsible for the exact solvability of the
Calogero's model.

This role of the Calogero's ${g}\,x^{-2}$ barriers is partially
weakened when they become attractive, i.e., $0 > g >-1/4$. An
extension of the Calogero model to this transition region has been
discovered and described by Gangopadhyaya and Sukhatme
\cite{Sukhatmeag}. In their construction the contact terms also
appeared as a formal means of preservation of the solvability. As
a consequence, the spectrum only consisted of the two shifted sets
of equally spaced energy levels. Such a type of the modified
Calogero's spectrum re-appeared also in our recent non-Hermitian
construction \cite{Milos}.

In the present letter we were able to get rid of the Calogero's
power-law ${\cal O}(x^{-2})$ barriers completely. Our alternative
way of introduction of the contact terms enabled us to treat their
couplings as independent parameters. Our key point is that these
barriers are ``thin", i.e., partially penetrable. This represents
their main phenomenological appeal. At all their finite (and, in
fact, both repulsive as well as attractive) couplings $G_j$, their
free variability might prove appealing in many phenomenological
considerations.

In contrast to the Calogero model characterized by the absolute
absence of tunneling, all our particles are permitted to jump over
one another. In certain applications to few body systems of quark
type, this could improve our intuitive insight and build some
analogies with the motion in more dimensions. After all, the use
of the harmonic two-body forces with an additional, contact
``local spike" might also extend the advantages of the exact
solvability quite easily beyond the traditional domains in the
theoretical nuclear physics.

\subsection*{Acknowledgement}

Work supported by the Czech GA AS, contracts No. A 1048004 and A
1048101.

\section*{Figure captions}

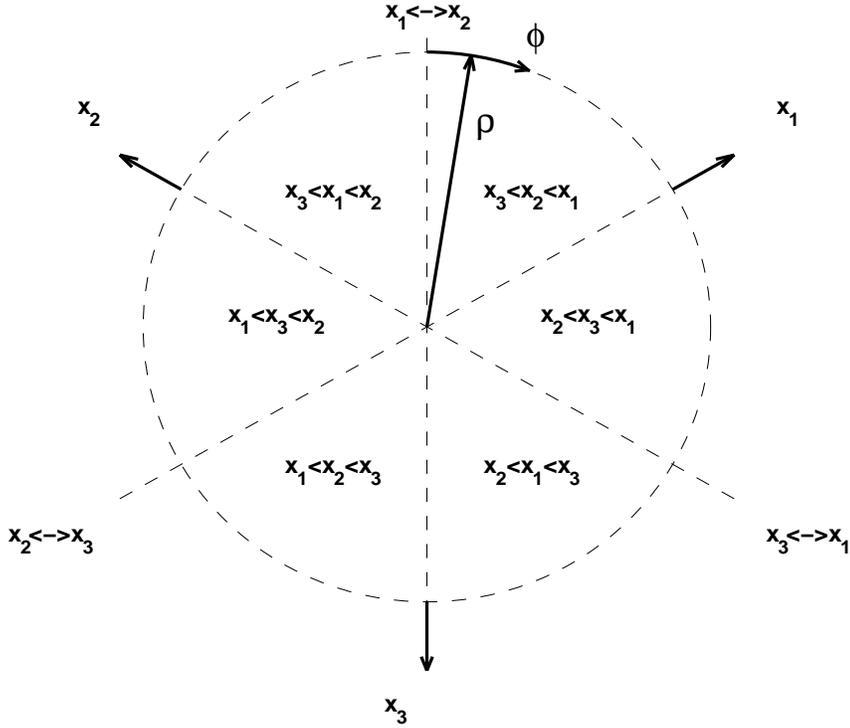
\subsection*{Figure 1. The choice of coordinates for three particles
}

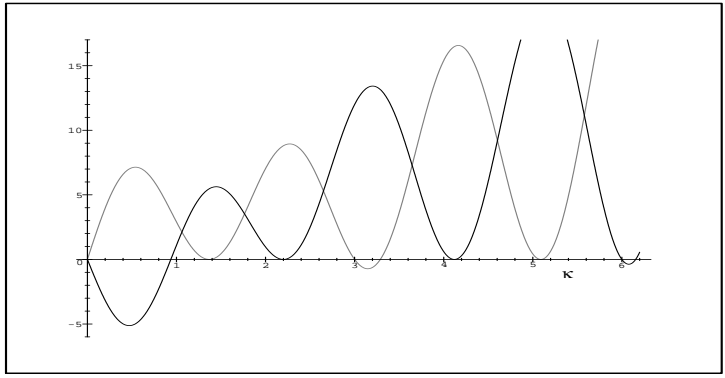
\subsection*{Figure 2. Graphical determination of the roots
$\kappa=\sqrt{E}$ at $G_1=G_2=G_3=1$ \label{brew}}

 \end{document}